\documentclass{article}

\usepackage{arxiv}

\usepackage[utf8]{inputenc} % allow utf-8 input
\usepackage[T1]{fontenc}    % use 8-bit T1 fonts
\usepackage{hyperref}       % hyperlinks
\usepackage{url}            % simple URL typesetting
\usepackage{booktabs}       % professional-quality tables
\usepackage{amsfonts}       % blackboard math symbols
\usepackage{nicefrac}       % compact symbols for 1/2, etc.
\usepackage{microtype}      % microtypography
\usepackage{lipsum}

\usepackage{graphicx}
\usepackage{amssymb}
\usepackage{amsmath}
\usepackage[flushleft]{threeparttable}

\title{Chunkflow: Distributed Hybrid Cloud Processing of Large 3D Images by Convolutional Nets}

\author{
  Jingpeng Wu\\
Princeton University\\
Princeton Neuroscience Institute, Princeton University, Princeton, NJ, USA\\
  \texttt{jingpeng@princeton.edu} \\
  \And
  William M. Silversmith\\
Princeton University\\
Princeton Neuroscience Institute, Princeton University, Princeton, NJ, USA\\
\textit{william.silversmith@princeton.edu} \\
    \And
    Kisuk Lee \\
    Massachusetts Institute of Technology\\
    Brain and Cognitive Science Department, Massachusetts Institute of Technology, Cambridge, MA 02139, USA\\
    \textit{kisuklee@mit.edu}\\
   \And
 H. Sebastian Seung\\
Princeton University\\
Princeton Neuroscience Institute, Princeton University, Princeton, NJ, USA\\
Department of Computer Science, Princeton University, Princeton, NJ, USA\\
  \texttt{sseung@princeton.edu} \\
}

\begin{document}
\maketitle

\begin{abstract}

It is now common to process volumetric biomedical images using 3D Convolutional Networks (ConvNets). This can be challenging for the teravoxel and even petavoxel images that are being acquired today by light or electron microscopy. Here we introduce chunkflow, a software framework for distributing ConvNet processing over local and cloud GPUs and CPUs. The image volume is divided into overlapping chunks, each chunk is processed by a ConvNet, and the results are blended together to yield the output image. The frontend submits ConvNet tasks to a cloud queue. The tasks are executed by local and cloud GPUs and CPUs. Thanks to the fault-tolerant architecture of Chunkflow, cost can be greatly reduced by utilizing cheap unstable cloud instances. Chunkflow currently supports PyTorch for GPUs and PZnet for CPUs. To illustrate its usage, a large 3D brain image from serial section electron microscopy was processed by a 3D ConvNet with a U-Net style architecture. Chunkflow provides some chunk operations for general use, and the operations can be composed flexibly in a command line interface.
\end{abstract}

% keywords can be removed

\section{Introduction}
Neurons can extend in the scale of centimeters and connect to others with its nanometer scale synapses. In order to map the connectivity of neurons, we need to perform high resolution imaging  with a large scale field of view. Using light \cite{migliori2018light} and electron microscopy \cite{morgan2016fuzzy}, neuroscientists are producing terascale or even petascale 3D images, and a great number of computer vision tasks, such as image segmentation, need to be performed to analyze the morphology and connectivity of neurons. Manual labeling is normally time consuming and tedious for large scale images, and automation using computational technologies is highly needed ~\cite{akil_challenges_2011}.

The number of parameters of a modern 3D ConvNet can be over 1 million ~\cite{kisuk_superhuman_2017} or even 83 million ~\cite{funke2018structured}. Consequently, the computation is expensive and time consuming for large 3D image datasets even with the acceleration of GPUs. Furthermore, RAM usage could be orders of magnitude of image size. As a result, traditional image processing with standalone software packages ~\cite{eliceiri_biological_2012} in a single workstation become insufficient to handle large scale image datasets. To accelerate the computation, distributed computation was needed to finish the computation in reasonable wall clock time. Some customized clusters were setup to perform distributed ConvNet inference ~\cite{kaynig_large_2015, plaza_large_2016, haehn_scalable_2017}, However, large local clusters were not widely accessible for a broader neuroscience community. For large scale ConvNet inference, the task is bursty meaning we only need a large number of machines in a short time, but local clusters normally have a fixed number of computers. It is a waste of resources to maintain a large computer cluster for occasional use.

Cloud computing platforms provide elastic computation as a service. The users can allocate the computational resources needed and just pay for what they really use. Big public cloud computing platforms, such as Google Cloud and Amazon Web Services (AWS), have enough computational power, storage capacity and bandwidth to handle terabyte or even petabyte scale of data. Users can launch a great number of instances to finish jobs in a short time, which is well suited for ConvNet inference of large scale image datasets. Furthermore, many cloud computing companies sell idle machines with a much lower price. Compared with the local usage of in-house cluster, the computational resources of the cloud are available globally and could be used by almost all of the neuroscientists. 

Currently, there exist some tools based on cloud to perform ConvNet inference for 2D images, such as Google Vision API and AWS Rekognition. The images are independent with each other and these services are not applicable for a complete 3D image dataset. Recently, Haberl Matthias \textit{et al}. setup a tool in AWS, called CDeep3M, to segment 2D/3D image stacks including training and inference, but it focus on single instance with small image stacks without distributed computation~\cite{haberl_cdeep3m_2018}. To our knowledge, there is a lack of a cloud computing tool to perform terabyte or petabyte scale 3D image dataset ConvNet inference with multiple frameworks utilizing both GPUs and CPUs. Here, we report a hybrid cloud framework, called chunkflow, to perform this task. Chunkflow was designed with hybrid cloud architecture to use both cloud and local computational resources, including both GPUs and CPUs, to maximize efficiency and reduce cost.

%-------------------------------------------------------------------------
\section{System Overview}

\subsection{Hybrid Cloud Infrastructure}
\begin{figure*}
\begin{center}
\includegraphics[width=.9\linewidth]{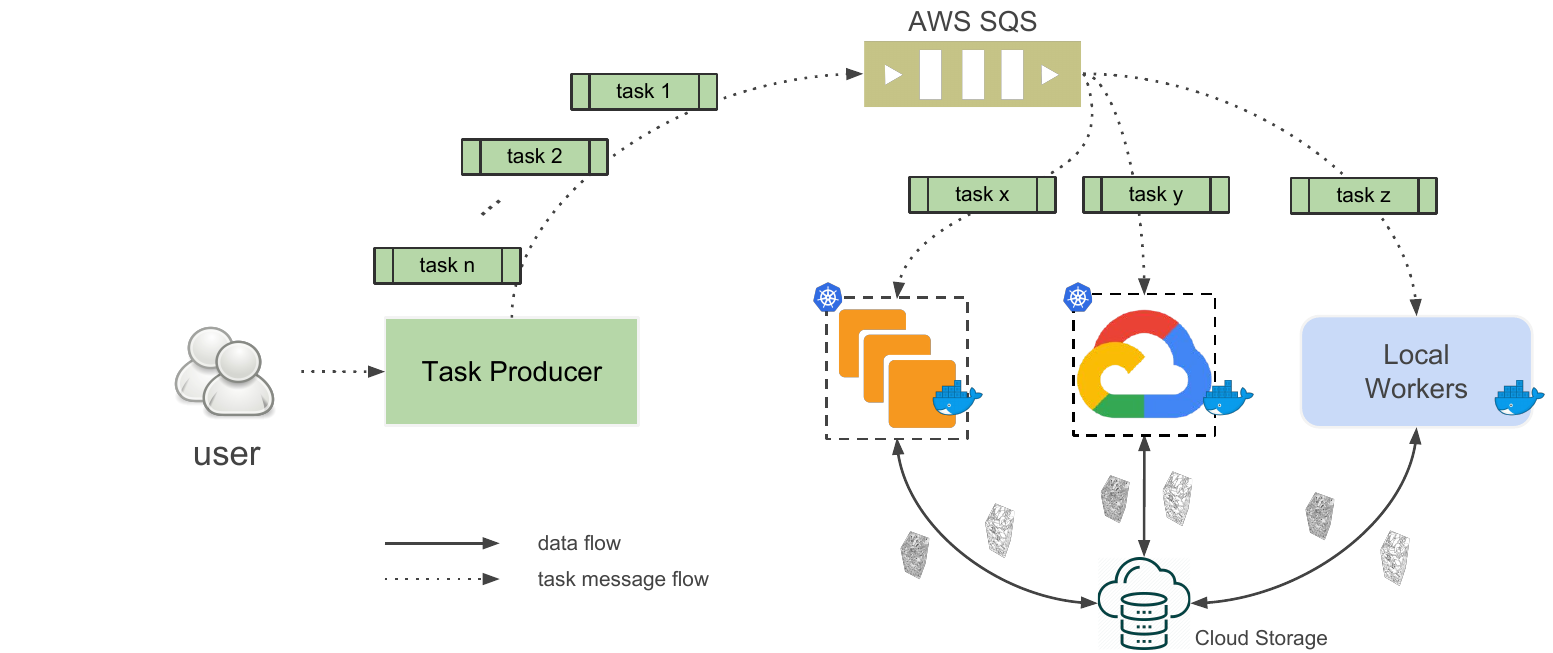}
\end{center}
\caption{Overview of chunkflow framework. User submits tasks to a queue in AWS SQS, the workers in cloud or local keep fetching tasks from the cloud queue and perform computation task. Both input and output chunks were stored in cloud storage. The workers in cloud could be managed using kubernetes.}
\label{fig:overview}
\end{figure*}

The whole 3D image dataset was chopped as overlapping chunks, and each chunk was distributed and processed as a task in a worker instance. As illustrated in Fig.~\ref{fig:overview}, chunkflow has a producer-consumer architecture. The task producing frontend and task consumption backend were decoupled using a queue in cloud. The user can produce a number of tasks specifying the output bounding box of each task. The tasks were generated and submitted to a message queue managed by the AWS Simple Queue Service (SQS). The users can also view, insert, edit or delete tasks directly using the web console of AWS SQS. In the computationally heavy backend, the workers first fetch task messages from the SQS queue, and perform the processing of chunk. After finishing the task, the workers will delete the task in queue and continue fetching new task to work on until there is no task left in the queue.

Benefited from the decoupling design and the high availability of AWS SQS, the computationally heavy workers in the backend could come from anywhere with internet connection and cloud authentication, including both the cloud and local computers. This design prevents vendor binding of computationally heavy backend and enables full utilization of available computational resources. For example, most labs have a few computers or a small cluster for algorithm development and small scale processing, which could also be fully utilized while production run. 

\subsection{Dataset Management}
The 3D image dataset was saved as non-overlaping blocks in cloud storage. The format follows the convention of Neuroglancer Precomputed for immediate visualization using neuroglancer~\cite{jeremy_webgl_2019}. Briefly, the blocks were saved as compressed binary files, and a JSON formatted file provides the metadata of the dataset including data type, number of channels, voxel size and volume size. The dataset could be downsampled and have multiple mip levels for visualization navigation. We use cloudvolume~\cite{silversman_cloudvolume_2019} for cutout and saving. Cloudvolume support multiple storage backends, including Google Cloud Storage, AWS Simple Storage Service, bossDB ~\cite{kleissas_block_2017}, and local filesystem. Although cutout could be arbitrary region, the saving region have to align with the block size in the storage backend to avoid overwriting from multiple workers. 

\subsection{Task Production and Management in Frontend}
Since we use the AWS SQS to manage tasks, producing tasks to SQS queue is an independent step without the setup of computation workers. Users can simply use the SQS web console directly for adding and editing the tasks or even develop their own task producer. The output volume space was decomposed as evenly spaced grids, and each task corresponds to a grid node. For the output cross the dataset boundary, the chunk will be automatically cropped by cloudvolume. 

% To eliminate the software installation and configuration overhead, we also built a simple web front-end. The user can just edit task template and change parameters, such as input and output paths. The web server will produce a serials of tasks to SQS accordingly. 

% \begin{figure}[h!]
% \begin{center}
% \includegraphics[width=8cm]{frontend}
% \end{center}\caption{Web frontend for users. The computation graph was used to define IO and computation.}\label{fig:frontend}
% \end{figure}

After pushing the tasks to queue, the tasks were managed and scheduled by AWS SQS. Thus, the user can even produce tasks a few days before using them. After a computational worker fetches the task, the task was not deleted immediately but set as invisible to other workers for a prefixed amount of time, called invisibility timeout. If the worker finished the task successfully, the worker will send a request to delete the task using the task handle. If the worker failed to finish the task due to some causes, such as a cloud instance termination, no deleting request was sent to the queue. AWS SQS will reset the task as visible to other workers after the invisibility timeout. Thus, the invisibility timeout should be longer than the computation time in the workers, otherwise multiple workers could do the same task. With the help of invisible time scheduling, the tasks were guaranteed to be finished with task failures and the system is robust to worker failures. 

\subsection{Distributed Computation in the Backend}
Since any computer with internet connection and cloud authentication could fetch tasks from queue, read and write to cloud storage, the computation could be distributed everywhere rather than traditional single cluster. Most of cloud computing company sell idle instances with a much lower price than normal on-demand instances to fully utilize their resources. For example, Google Cloud sell these instances at a fixed lower price, called preemptible instances. Take the NVIDIA Tesla K80 GPU (not including other parts, such as CPU and RAM) for example, the preemptible price is $\$0.135$ per GPU compared with $\$0.45$ per GPU in on-demand instance (region us-central1), over three times of difference. AWS also sell these idle instances called spot instances, but they sell them in a free market with dynamically generated price. The users bid to compete with each other, and were only charged by the dynamic price lower than the bidding. Take the p2.xlarge instance with K80 GPU for example, the on-demand price is $\$0.90$ per hour in the us-east-1 region, and the average spot price is about $\$0.3$ per hour. While enjoying lower price, these instances could be forcefully terminated in favour of on-demand requests. Thus, utilizing these idle instances requires the application robust to instance termination. Since our chunkflow architecture was designed to be fault-tolerant, we can use these instances safely with much lower price. 

\subsection{Composable Command Line Interface for Flexible General Usage}
\begin{figure*}
\begin{center}
\includegraphics[width=.7\linewidth]{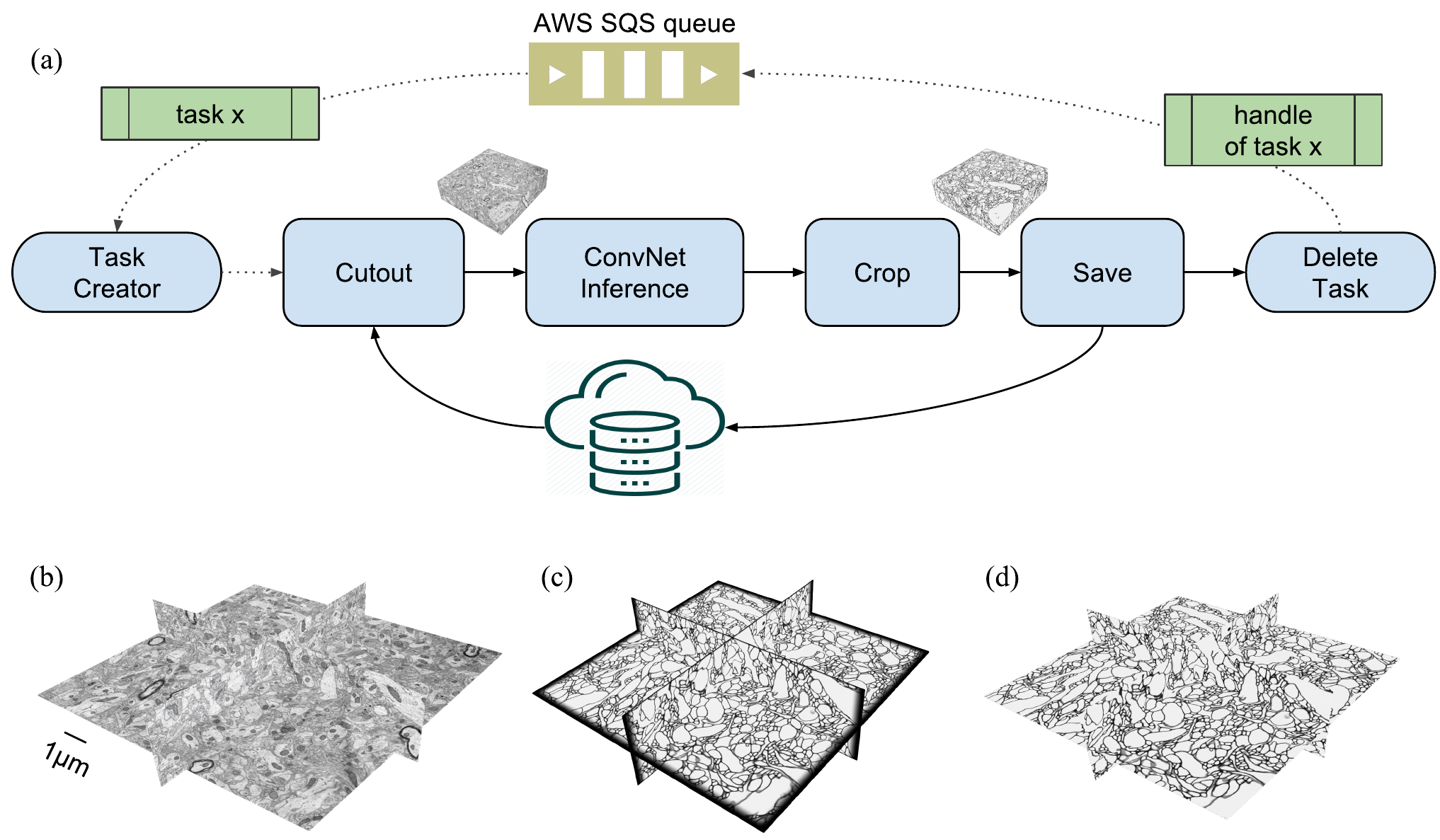}
\end{center}
\caption{A minimal example of composed pipeline for ConvNet inference of image chunks. (a) Structure of the example pipeline. (b) The image chunk cutout from the cloud storage. (c) The affinity map after inference of the image chunk. (d) The affinity map after cropping the chunk margin.}
\label{fig:composed_pipeline}
\end{figure*}

According to our production run experiences, it turns out hard to build a universal pipeline for all the serial section electron microscopy image datasets and we normally need to adjust the pipeline for each dataset. For example, the duplicate image section might need to be masked out or the affinity map needs to be masked out due to misalignment. Thus, we built a composable command line interface to flexibly stitch the chunk operators in production. The computation in each computational node was abstracted as three types: Chunk Generator, Chunk Processor and Utility. Task Generator is responsible to generate new tasks. For example, one of the generator can keep fetching task from AWS SQS queue and generate a serials of tasks. Chunk Processor is responsible to process the chunk, such as masking, ConvNet inference, and cropping. Utility is responsible to do some assistant jobs, such as uploading logs to cloud storage, ingesting processor time elapse to AWS CloudWatch as a real-time speedometer, and deleting the finished task in AWS SQS queue. Fig.~\ref{fig:composed_pipeline} illustrates a minimal example of composed pipeline for ConvNet inference. Currently, we have 12 operators implemented and more will be added in future. All the operators have user friendly help message to guide the usage. 

\subsection{ConvNet Inference using Both GPU and CPU}
Most of ConvNet training frameworks were heavily optimized for GPUs, but GPUs are not always available due to the heavy usage in training. According to our production runs in the cloud, it is sometimes hard to allocate cheap GPU instances in big public cloud vendors, such as AWS and Google Cloud. Luckily, well optimized network could be cheaper to run in CPU than GPU~\cite{zlateski_compile_2017, popovych_pznet_2019}. Thus, chunkflow currently support both GPU and CPU for ConvNet inference. 

For inference using GPU, we currently support PyTorch and more will be added in the future. For inference using CPU, we compile the ConvNet model with PZnet~\cite{popovych_pznet_2019}, a python wrapper of znnphi ~\cite{zlateski_compile_2017}. Basically, PZnet optimize the computation at the whole network level rather than layer level and optimize the cache efficiency automatically according to the cache size of workers. The inference computation was decomposed to small tasks and the tasks were scheduled statically. For each network model, we need to compile once for a fixed number of cores. Then, we perform ConvNet inference using the compiled convnet. 

\begin{figure}[t]
\begin{center}
\includegraphics[width=0.5\linewidth]{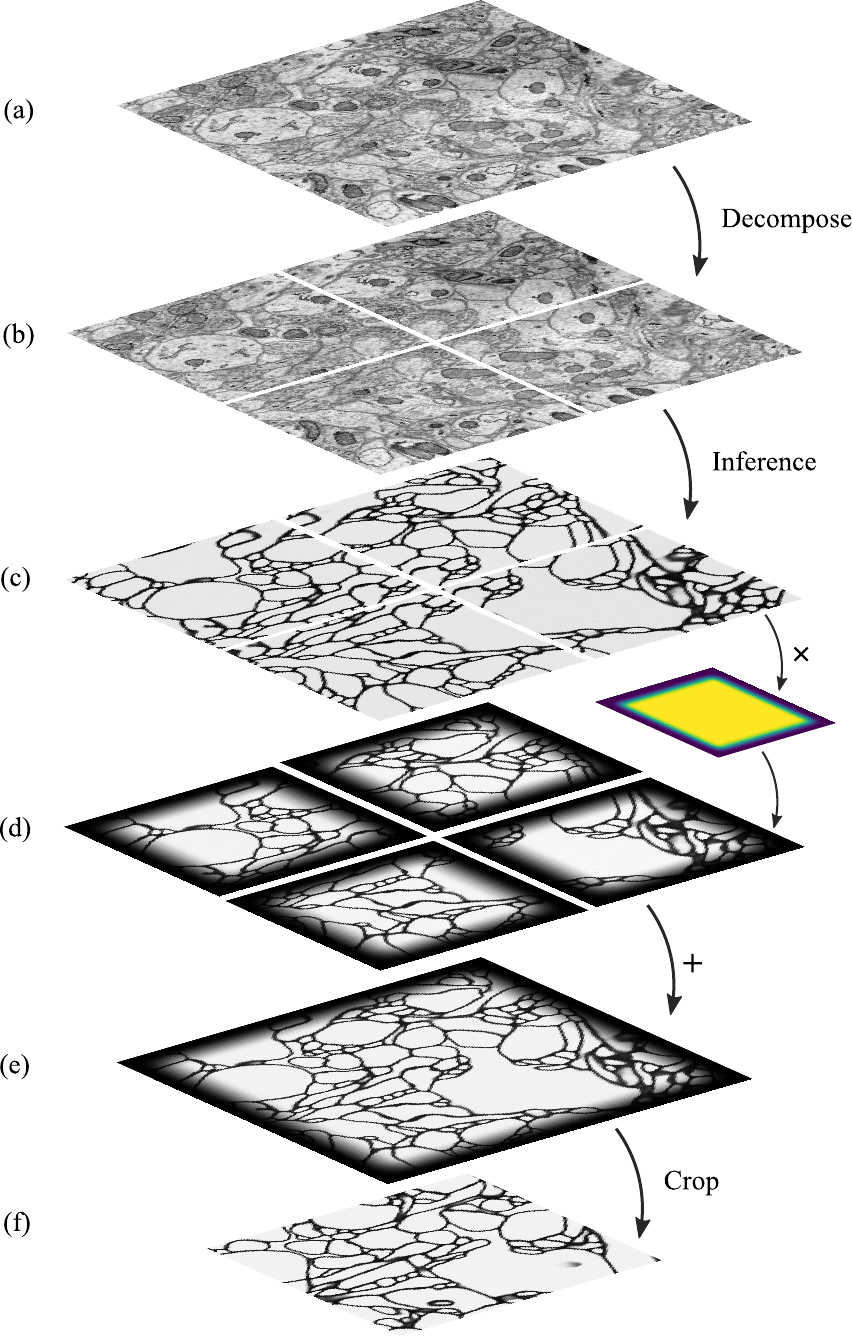}
\end{center}
\caption{ConvNet output patch normalization for overlapping regions. The raw image chunk (a) was decomposed to overlapping patches, and the patches was passed through a ConvNet to produce affinity map (b), the affinity map voxels were weighted according to the distance to boundary (d), the the weighted affinity map was recombined by summation to form a larger map (e), the larger map was then cropped to remove the weighted margin (f). Note that the process was illustrated in 2D for simplicity, and the actual processing is in 3D. This illustration used $2 \times 2$ patches, but actual processing have much larger number of patches in 3D.}
\label{fig:patch_mask}
\end{figure}

Mordern ConvNet normally have millions of parameters and could use as much as three orders of magnitude of RAM during inference. As a result, the image chunk can not fit in the memory with a typical instance/machine. Thus, we have to decompose the image chunk to small patches (Fig. ~\ref{fig:patch_mask} a,b) and perform inference patch by patch (Fig.~\ref{fig:patch_mask} c). To reduce boundary effect among neighbouring patches, we make the patches overlap with each other and the overlapping region was blended. The overlapping region of neighboring patches were weighted using a bump function(Equation~\ref{eq:bump}), and normalized according to the weight (Fig.~\ref{fig:patch_mask} d). The weight of each voxel $\omega(d)$ was computed according to the normalized distance $d$ from the center of the patch. The center of patch have higher weight since it has less boundary effect than the voxels around the patch border. The weights were accumulated around the overlapping region and renormalized to make sure that the summation of weights in overlapping regions equals to one. The normalization mask only depends on the patch size and overlap size, and could be precomputed to reduce blending overhead. The mask was applied to every patch directly, and only element wise summation was needed to blend the overlapping patches. 
\begin{equation}
\omega(d) = 
\begin{cases}
\exp \left( -\frac{1}{1-d^2} \right) & \mbox{ for } d\in (-1, 1) \\
0 & \mbox{ otherwise }
\end{cases}
\label{eq:bump}
\end{equation}

After blending the patches, we get a output chunk, but the output chunk also have boundary effect if stitched together directly. Thus, we cutout image chunks with overlap and crop the marginal regions of the output chunk (Fig.~\ref{fig:composed_pipeline} d, Fig.~\ref{fig:patch_mask} f). The cropped output chunk was compressed and uploaded. The output normally have a data type of single precision floating point with multiple channels, and the size is much larger than image with bitdepth of 8. As a result, the block size is relatively big and loading in neuroglancer could be slow if the network bandwidth is not large enough. To visualize the output faster, chunkflow have an option to quantize and downsample the output, called thumbnail, for fast visualization.

\subsection{Utilizing Cheap Cloud Instances}
Benefited from the fault-tolerant architecture, we can use cheap idle instances in the cloud, such as preemptible instance in Google Cloud and spot instance in AWS. The cheap instances are not stable, workers could be terminated in need of the cloud vendors. For example, AWS spot instance will be terminated if the bidding is lower than the market price. Thus, the framework must be robust to worker termination and other errors. In spite of internal retry with error handling, chunkflow also takes advantage of an AWS SQS feature to perform auto retry implicitly in the workers level. As a result, chunkflow can safely utilize these unstable cheap instances.

%Chunkflow was setup with continuous integration with auto-built docker image, so it could be deployed easily in different computational environments.

\section{Experiment}
\subsection{Dataset}

Chunkflow is an in-house production system with hundreds of terabyte scale runs for multiple large scale datasets. To illustrate the usage here, we performed ConvNet inference for a public dataset ~\cite{kasthuri_saturated_2015}. Briefly, a sample from mouse somatosensory cortex was sectioned with a thickness of 29nm and imaged using tape-collecting ultramicrotome and scanning electron microscope. The total number of coronal sections is 2250. The image ataset is publically available in Open Connectome Project ~\cite{burns_open_2013}. The original dataset was oversampled to 3 nm pixel size, so we downsampled back to 6 nm, and the size is $12286 \times 11262 \times 2046$. The dataset was also realigned for better automatic 3D segmentation using Alembic~\cite{macrina2019alembic}. The realigned dataset was ingested to Google Cloud Storage following the precomputed format of Neuroglancer ~\cite{jeremy_webgl_2019} with multiple mip levels. Our net was trained in mip 1 with voxel size of $12 \times 12 \times 29 $nm and size of $6143 \times 5631 \times 2046$. 

\subsection{ConvNet Training}
\begin{figure*}
\begin{center}
\includegraphics[width=0.8\linewidth]{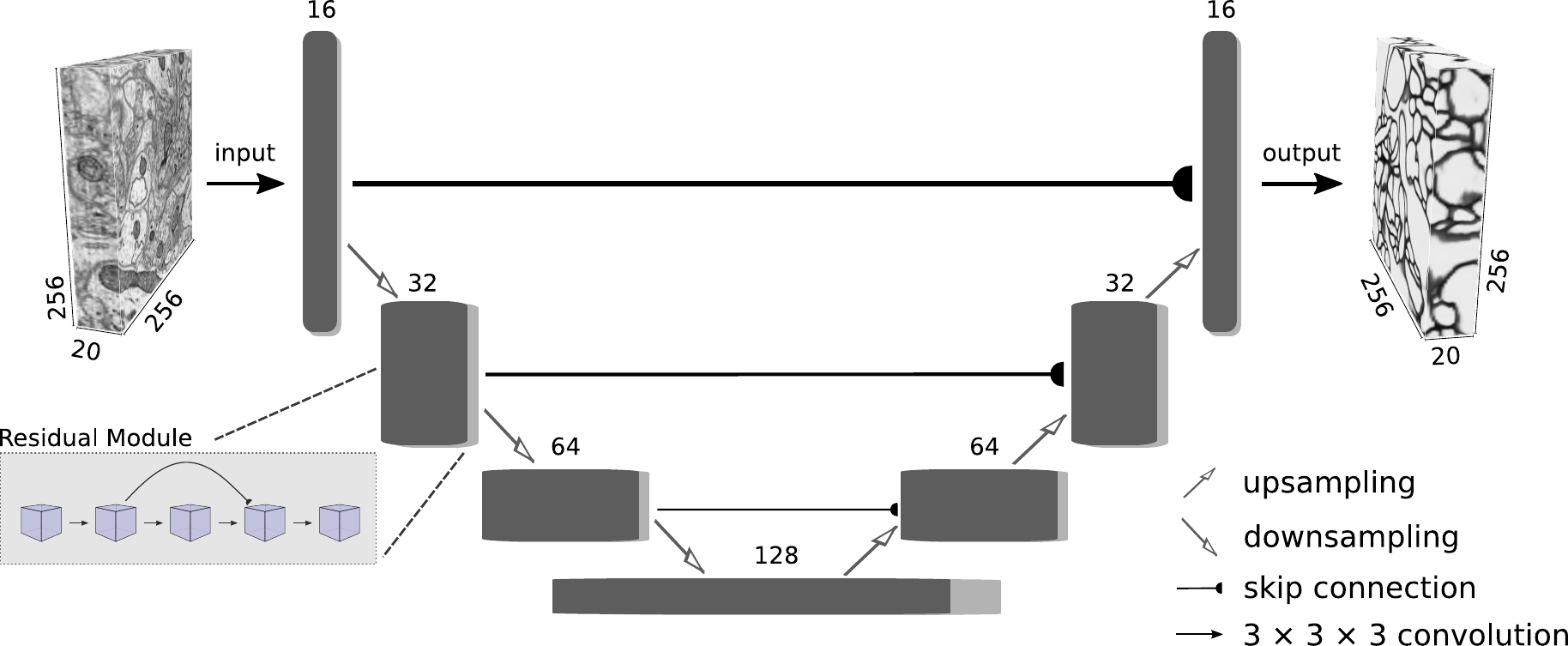}
\end{center}
\caption{ConvNet architecture.}
\label{fig:convnet}
\end{figure*}

The ConvNet training basically followed a leading method in the SNEMI3D challenge ~\cite{kisuk_superhuman_2017} which was built upon U-Net ~\cite{ronneberger2015u}. The network architecture was illustrated in Fig.~\ref{fig:convnet} with some modifications compared with ~\cite{kisuk_superhuman_2017}: (1) we used instance norm instead of batch norm; (2) we used all $3 \times 3 \times 3$ convolutions, except for the $5 \times 5 \times 1$ input/output convolutions; (3) we used ReLU rather than ELU; (4) we used bilinear resize convolution for upsampling instead of strided convolution transpose.

\subsection{Distributed Inference}
The tasks were produced and ingested to an AWS SQS queue. To demonstrate the hybrid cloud architecture, we used some computational instances from both local and public cloud vendor, and both GPUs and CPUs. The local computation was performed using a workstation with a NVIDIA GeForce GTX 970 GPU device. In Google Cloud, we used 2 n1-standard-4 preemptable instances with 4 cores and 15 gigabyte of RAM for CPU inference using PZnet, and 2 n1-highmem-4 preemptable instances with 4 cores and 26 gigabyte of RAM with NVIDIA Tesla T4 GPU using PyTorch (2 worker processes in each instance). In AWS, we used one p2.xlarge spot instance (4 worker processes in each instance) with NVIDIA Tesla K80 GPU. The code was packaged in docker image, and the docker containers were deployed and managed using kubernetes in the cloud. 

As shown in Table~\ref{tab:performance}, Google instance with NVIDIA Tesla T4 GPU have the best performance and also good cost efficiency\footnote{The speed was measured with thousand voxel per second and the cost was measured with US dollar per billion output voxel. Note that we used CUDA9.1 and PyTorch 0.4.4 in AWS and CUDA10 and PyTorch1.0 in Google Cloud due to different kubernetes configuration. The AWS spot instance price was based on the price during the experimental run. The dataset was hosted in Google Cloud Storage, so AWS computer nodes took more time to perform IO.}. Note that the instance norm in ConvNet prevented the layer merging optimization in PZnet and made PZnet slower than network with batch norm. The total cost of inference was mainly composed by computation and storage. SQS message is free for 1 million requests per month, which is big enough in inference application. The cost of storage depends on the time kept in cloud, the data could be moved to archiving storage to reduce cost. 

% TO-DO add aws experiment
\begin{table}
\centering
%\begin{threeparttable}
\caption{Performance of computationally heavy backends.}
\begin{tabular}{lllll}
\toprule
Device 		& Framework	& Cloud 	& Speed 	& Cost	\\
\midrule
T4 			& PyTorch	& Google 	& 880 		& 0.11	\\
CPU 		& PZNet		& Google 	& 105		& 0.11	\\
K80			& PyTorch 	& AWS		& 679		& 0.14 	\\
GTX 970 	& PyTorch 	& Local 	& 471 		& --	\\
\bottomrule
\end{tabular}
%\begin{tablenotes}
%\small
%\item The speed was measured with thousand voxel per second and the cost was measured with US dollar per billion output voxel. Note that we used CUDA9.1 and PyTorch 0.4.4 in AWS and CUDA10 and PyTorch1.0 in Google Cloud due to different kubernetes configuration. The AWS spot instance price was based on the price during the experimental run. The dataset was hosted in Google Cloud Storage, so AWS computer nodes took more time to perform IO.
%\end{tablenotes}
%\end{threeparttable}
\label{tab:performance}
\end{table}

\begin{figure}[ht!]
\begin{center}
\includegraphics[width=0.6\linewidth]{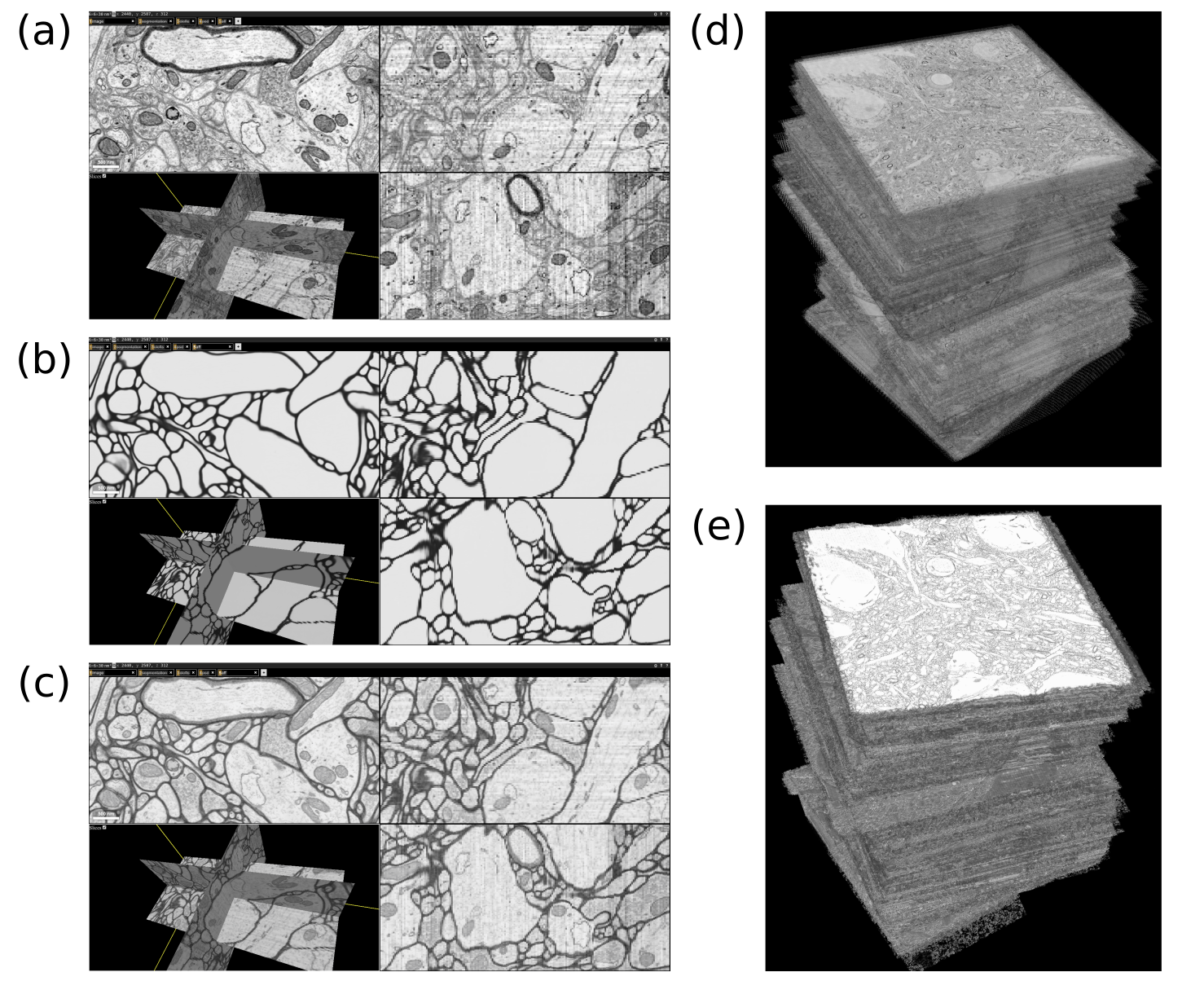}
\end{center}
\caption{Visualization of ConvNet inference result. The result could be visualized in real time in neuroglancer including image (a), first channel of affinity map (b) and combined (c). The whole volume of image and affinity map were also rendered after downsampling.}
\label{fig:image_aff}
\end{figure}

The resulting affinity map is about 790 gigabyte with a data type of single precision floating point and 3 channels. The channels represent the afffinity value between neighboring voxels in the X,Y,Z direction ~\cite{briggman_maximin_2009}. The downsampled volume of image and quantized affinity map was rendered to illustrate the whole dataset(Fig.~\ref{fig:image_aff} d,e).

\section{Discussion}
\subsection{Architecture Design}
Traditional solution of large scale ConvNet inference is building a large cluster ~\cite{plaza_large_2016}. Given the fast evolution of deep learning software and hardware, it could be technically hard and wasteful for the adoption of other labs for occasional use. Currently, most of labs just build their own pipelines for these kinds of tasks with few reports of infrastructure reuse across labs. Chunkflow was designed to have a hybrid cloud architecture with flexible deployment. It does not have vendor or cluster binding for the computationally heavy backend. Benefited from the global infrastructure of Google and AWS, chunkflow could be deployed around the world and have much higher availability than cluster centred setup. Chunkflow was distributed using docker image with continuous integration, so we can keep the users updated seamlessly. The fault-tolerant design enables us to use cheap instances of public clouds, such as preemptable instances in Google Cloud and spot instances in AWS. 

Since GPU is popular in ConvNet training, the availability in public cloud could be low. The GPU spot instances in AWS is so popular that the bidding price can keep ten times of on-demand instance price for days, which indicates that all the GPUs were fully used as on-demand instances and are not available in the bidding market. In contrast, CPU instances have much higher availability with much cheaper spot price. For some ConvNet, the highly efficient PZnet can make CPU inference cheaper than GPU instances in terms of throughput per dollar ~\cite{zlateski_compile_2017, popovych_pznet_2019}. 

Since we make data format consistent with neuroglancer precomputed, the affinity map could be visualized in real time while we run inference (Fig.~\ref{fig:image_aff} a-c). Users can check the quality and correctness while processing, which is helpful and important to diagnose problems in an early stage, especially for large scale image processing. Processing large scale image dataset with bug would be a big waste of resources. Currently, most of quantitative evaluation methods were built upon manually labelled groundtruth, which is normally limited in scale, and large scale evaluation is still relying on visual inspection of results. With the direct visualization of processing result, especially the quantized and downsampled thumbnail, users can also easily performing large scale evaluation. According to our production runs, although the ConvNet was trained with a lot of ground truth label, it is still hard to generalize well across large scale image dataset due to a great variety of brain structures. Being able to perform large scale evaluation and picking out problematic regions is useful for improving model generalization, such as focused annotation and training.

\subsection{Limitations and Future Directions}
To reduce the boundary effect of each chunk, the inference result was cropped around chunk boundary. Theoretically, the result across chunk boundary is globally consistent if the cropping size is half of patch size, but it would be a big waste of computation in practice and we normally use cropping size less than half of patch size. As a result, there is a balance between consistency and efficiency. Although the inconsistency is normally hard to notice with small cropping size (Fig.~\ref{fig:image_aff}), it would be better to eliminate it in the first place. One solution is reusing the chunk margins by introducing task dependency, which will require a major architecture change. Reusing the cropped chunk margin would also reduce the computational cost since less number of patches was required with the cost of increasing IO and compression/decompression overhead. In practice, it depends on the resource availability to choose approaches.

Although the inference using CPUs with PZnet was highly optimized, the inference using GPUs with PyTorch was not well optimized yet. TorchScript could be used to serialize and optimize the model in the future. Benefited from the inference backend design, other frameworks with compilation and optimization could also be added easily. To add a new inference backend, we only need to implement 5D patch processing including batch size, number of channel and patch size. 

\section*{Conclusion}
In summary, chunkflow is a hybrid-cloud framework for distributed ConvNet inference of large scale 3D image datasets. Due to the decoupled design of task production and consumption, almost all the computers with internet connection and cloud authentication could be used as computational backends. Benefited from the robust design to handle task failure, chunkflow could utilize the idle instances in the cloud to greatly reduce the computational cost. The composable commandline interface also makes the general usage flexible. Since chunkflow was built around public cloud services without vendor binding in backend, it could be deployed around the world to perform distributed 3D ConvNet inference.

\bibliographystyle{unsrt}  
\bibliography{references} 
\end{document}